\documentstyle[aps,multicol,epsfig,prl,floats]{revtex}

\begin{document}
\draft
\twocolumn[\hsize\textwidth\columnwidth\hsize\csname
@twocolumnfalse\endcsname
\title{Impurity and strain effects on the magnetotransport of
La$_{1.85}$Sr$_{0.15}$Cu$_{1-y}$Zn$_y$O$_4$ films}
\author{Marta Z. Cieplak,$^{1,2}$ A. Malinowski,$^{1,2}$ K.
Karpi\'{n}ska,$^2$ S. Guha,$^1$ A. Krickser,$^1$ \\
B. Kim, $^1$ Q. Wu, $^1$ C. H. Shang, $^3$ M. Berkowski,$^2$ and P. Lindenfeld$^1$}
\address{$^1$ Department of Physics and Astronomy, Rutgers University,
Piscataway,
NJ 08855, USA\\
$^2$ Institute of Physics, Polish Academy of Sciences, 02 668 Warsaw,
Poland\\
$^3$ Department of Physics and Astronomy, The Johns Hopkins University,
Baltimore, Md 21218, USA}
\maketitle

\begin{abstract}
The influence of zinc doping and strain-related effects on the normal
state transport properties (the resistivity, $\rho$, the Hall angle,
$\Theta_H$, and the orbital magnetoresistance, $\Delta \rho /\rho$)
is studied in a series of La$_{1.85}$Sr$_{0.15}$Cu$_{1-y}$Zn$_y$O$_4$
films, with values of $y$ between zero and 0.12 and various degrees of strain
induced by the mismatch between the films and the substrate.
The zinc doping affects only the constant term in the temperature
dependence of $\cot {\Theta _{H}}$ but the strain affects both the
slope and the constant term, while their ratio remains constant.
$\Delta \rho /\rho$ is decreased by zinc doping but is unaffected
by strain. The ratio $({\Delta}{\rho}/{\rho})/{\tan ^2{\Theta _{H}}}$ is
$T$-independent but decreases with impurity doping. These results put 
strong constraints on theories of the normal state of high--temperature 
superconductors.
\end{abstract}


]

The normal state of the high--$T_c$ superconductors is usually called
{\it anomalous} because its properties differ from those of other
metals. In this letter we discuss the resistivity, $\rho$, the Hall
angle, $\Theta_H$, and the orbital magnetoresistance (OMR),
$\Delta\rho/\rho$ . Within well--known limits these quantities
have been shown to follow the relations $\rho = \rho_0 +
AT$, $\cot{\Theta_H} = \alpha T^2 + C$, and $\Delta\rho/\rho = \zeta
\tan^2{\Theta_H}$ \cite{gurv,chien,harris,fedor}.

We have investigated these properties in films of
La$_{1.85}$Sr$_{0.15}$Cu$_{1-y}$Zn$_y$O$_4$ (LSCO) with varying amounts
of strain as evidenced by different values of the $c$-axis lattice parameter, 
$c$, and with different amounts of zinc
substituted for the copper. The strain and the zinc impurities affect
the properties differently, and we have been able to modify the earlier
equations so as to separate the results of the two effects. The least
studied parameter is $\zeta$, and we show that it has a surprisingly simple
impurity dependence. The
experiments also show that while the Hall effect is strongly influenced
by strain, the OMR is not. This unexpected result places strong
constraints on theories of the anomalous normal state.

The specimens were made by pulsed laser deposition on substrates 
of LaSrAlO$_4$ \cite{cieplak1}. They are
$c$--axis oriented, about 6000 {\AA} thick. The values of the
zinc--fraction, $y$, (from 0 to 0.12) are those of the targets, but have
been shown to be the same as those of the films \cite{cieplak2}. The
details of the specimen preparation and of the  measurements have been
described previously \cite{fedor,cieplak2}. The parameters $\rho_0$ and
$A$ were determined from the linear dependence on temperature of $\rho$ between
200K and 300K, and the parameters $\alpha$ and $C$ from the quadratic
dependence on temperature of $\cot {\Theta_H}$ between 70K and 200K at 8
tesla.

For any given value of $y$ the films grow with randomly varying amounts
of built-in strain, which we utilize to study the effect of strain
on the transport. We first discuss the origin of the strain variation
for the films with $y = 0$.

\begin{figure}[ht]
\vspace*{-0.8cm}
\epsfig{file=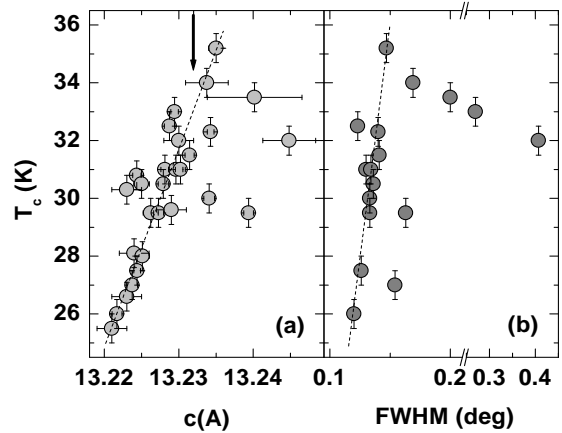, height=0.66\textwidth, width=0.47\textwidth}
\vspace*{-4.7cm}
\caption{$T_c$ as a function of $c$-axis lattice parameter (a) and 
$T_c$ as a function of FWHM (b) for a series of LSCO films without 
zinc. The arrow in (a) indicates the value of bulk lattice parameter.
The dotted lines are fitted to the data points for the films with
small FWHM.}
\end{figure}

\begin{figure}[ht]
\vspace*{-1.0cm}
\epsfig{file=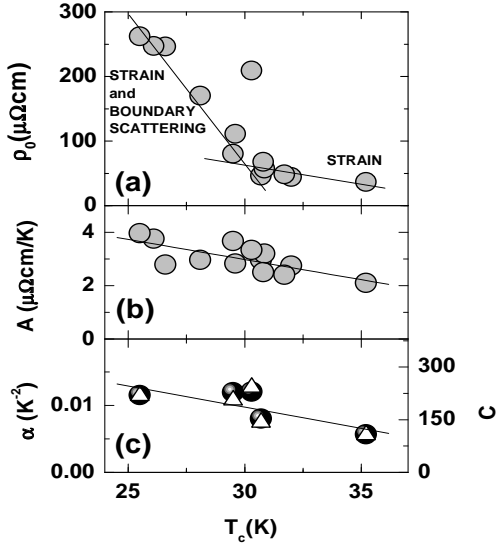, height=0.66\textwidth, width=0.50\textwidth}
\vspace*{-3.4cm}
\caption{Transport parameters $\rho_0$ (a), $A$ (b), $\alpha$ (full
circles) and $C$ (open triangles)(c) as a function of $T_c$ for a series
of LSCO films without zinc. The straight lines show linear fits to the 
experimental data. The definitions of all transport parameters are given in the
text.}
\end{figure}

The correlation between the superconducting transition temperature, 
$T_c$, and the $c$-axis lattice parameter, $c$, is shown on Fig.~1a 
for films with $y$ = 0, and is similar to that of the films of Sato 
{\it et al.} \cite{sato2} with varying La--Sr ratios.
The lattice parameters were determined from least-square fits to
eight high-angle $(00l)$ peaks, and the error bars are the standard
deviations from the fits. In several films the error bars are large,
reflecting real distributions of $c$--values in these films. Fig.~1b 
shows $T_c$ versus the full width at half maximum (FWHM) of rocking
curves measured for the (008) peak. The films with a large $c$--distribution 
also have large FWHM. In most of the remaining films
FWHM is below 0.15 deg indicating good crystalline quality. For these 
films $T_c$ is proportional to $c$, and also to FWHM, as shown by 
the dashed lines fitted to the data in both panels of Fig.~1.

The substrate's $a$--axis parameter is 3.756 {\AA}, compared to the bulk
value for LSCO of 3.777 {\AA}, resulting in compressive in--plane
mismatch, which should be accompanied by an expansion of the $c$--values.
However, the films grow with $c$ either expanded or compressed with
respect to the bulk value (13.232 {\AA}), implying either compressive or
tensile in-plane strain. Indeed, using a 4--cycle diffractometer we have verified
for several films that this is the case. The strain, defined as
$\epsilon_d =(d_{bulk}-d_{film})/d_{bulk}$ (where $d$ is the
lattice parameter), ranges from about $\epsilon_a = +0.05$\% and
$\epsilon_c = -0.02$\% (compressive in--plane strain) for the film with
the highest $T_c$, to about $\epsilon_a = -0.19$\% and $\epsilon_c =
+0.08$\% (tensile in--plane strain) for the film with the lowest $T_c$.
Using uniaxial strain coefficients from Ref. \cite{gugen} we estimate that
the change of $T_c$ by compressive or tensile strain in these two
extremes should be of the order of 1 K and $-2$ K, respectively. 
In the region of compressive strain $T_c$ is relatively large, in keeping 
with the finding of Refs. \cite{sato2,sato1,locquet} that compressive 
strain enhances $T_c$. The suppression of $T_c$ by the tensile strain is 
somewhat larger than expected.

\begin{figure}[ht]
\vspace*{-1.0cm}
\epsfig{file=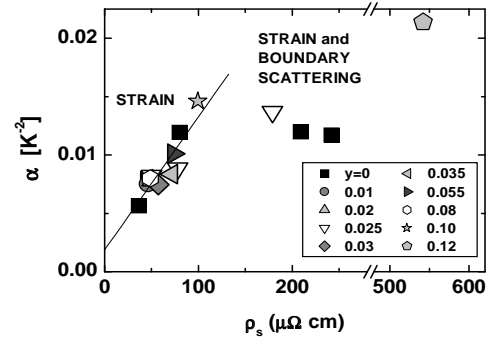, height=0.75\textwidth, width=0.55\textwidth}
\vspace*{-7.5cm}
\caption{
The dependence of $\alpha$ on the strain-related residual 
resistivity $\rho_s$. In the regime of $\rho_s < 100\, {\mu}{\Omega}\,$cm
the data fall on a straight line independent of zinc content. 
The presence of enhanced grain boundary scattering results in the deviation 
from this line for films with $\rho_s > 100\, {\mu}{\Omega}\,$cm.}
\end{figure}

In heteroepitaxial growth strain can either be relieved gradually as
the distance from the substrate increases, or by dislocations right at the
interface \cite{disloc}. A large $c$--distribution (large error bars in 
Fig.~1a) indicates gradual strain relief. Atomic Force Microscopy (AFM) 
shows that the growth of these films proceeds with grains of various sizes, 
some quite large, typical of 3D--growth. The root-mean-square (rms) roughness 
reaches as high as 7 nm.

In most of the remaining films these features are absent, suggesting that 
dislocations at the interface relieve the strain. FWHM decreases in the films 
with tensile strain and low $T_c$. Therefore random disorder (such as from 
oxygen vacancies) cannot explain the suppression of $T_c$. We have examined 
AFM images for films along the dashed line in Fig.~1. The films with $T_c$ 
higher then 30 K (where the strain is negligibly small or compressive) show 
very smooth surfaces with very well fused grains and a rms roughness as small 
as one unit cell. As $c$ departs more and more from the bulk value and $T_c$ 
drops below 28 or 30 K, the rms roughness increases. In films with the lowest 
$T_c$ the grains are very flat but with substantial discontinuities between 
them, and the rms roughness increases again to 6 nm. We interpret this behavior 
as indicating an increasing density of dislocations at the interface as 
$T_c$ decreases. A small dislocation density leads to partial relief of 
strain at the interface, allowing the films to grow with compressive or 
negligible built-in strain and relatively high $T_c$. The increase of the
density of dislocations leads gradually to substantial imperfections at the 
grain boundaries, causing tensile strain in the grains and lower $T_c$.
The details of the structural studies will be published separately.

\begin{figure}[ht]
\vspace*{-1.0cm}
\epsfig{file=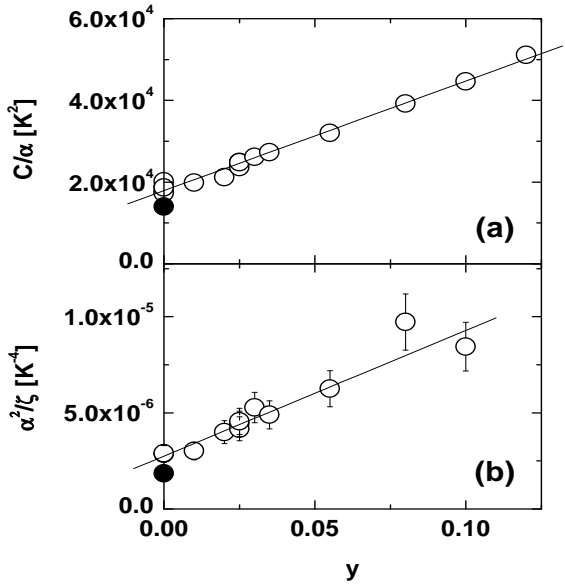, height=0.65\textwidth, width=0.5\textwidth}
\vspace*{-3.0cm}
\caption{
The ratio $C/{\alpha}$ (a) and ${\alpha ^2}/{\zeta}$ (b) versus zinc
content. The full circles are for the single crystal of Harris {\it
et al.}[3].}
\end{figure}

Fig.~2 shows the dependence of $\rho_0, A, \alpha$, and $C$ on $T_c$
for the films with $y = 0$. We see that the data for $A,
\alpha$, and $C$ fall on parallel lines, with fractional changes of
about 8.3\% per kelvin. For $\rho_0$ less than about $100\,\mu\Omega\,$cm
its dependence is also similar, but at higher values it changes more
rapidly. The crossover occurs when $T_c$ drops below 28-30 K, which is
the region where AFM images show increasing rms roughness.
This is consistent with an increasing density of dislocations
which produce imperfect grain boundaries and contribute to the
increased scattering. The grain boundary scattering apparently does not
influence other transport parameters besides $\rho_0$. The increase of
the parameter $\alpha$ confirms that the suppression of $T_c$ in the
tensile region is not related to a decrease of oxygen. A decrease of
oxygen would lead to a decrease of the carrier concentration,
and it is known that this leads to a decrease of the slope of the
cotangent \cite{fedor,hwang}, contrary to our results.

We now turn to the samples in which some of the copper is replaced by
zinc. For each value of $y$ the films again have different values of
$\rho_0$. The crossover in the dependence of $\rho_0$ on $T_c$, shown on
Fig.~2a for $y = 0$, persists for other values of $y$, as already reported in
Ref. \cite{cieplak2}. The minimum value of $\rho_0$ ($\equiv \rho_y$) 
for each $y$ is proportional to $y$, increasing at a rate 
$\rho_y /y = 2.76\,m{\Omega}\,cm$. The additional amount, 
$\rho_0 - \rho_y = \rho_s$, is then the part that depends on
strain and grain--boundary scattering.

Fig.~3 shows $\alpha$, the slope of the Hall--angle line, as a function
of $\rho_s$ for specimens with a range of values of $y$. We see a
linear part for small values of $\rho_s$, and, for 
$\rho_s > 100\,{\mu}{\Omega}\, cm$, a crossover to the region with 
increased (grain--boundary) scattering. We also see that $\alpha$ is a 
function of $\rho_s$, but not of $y$, that is, {\em the slope of the cotangent
line depends on strain, but is independent of impurity content}. 
This result indicates the validity of our separation of $\rho_0$ into 
$\rho_y$ and $\rho_s$. 

\begin{figure}[ht]
\vspace*{-.7cm}
\epsfig{file=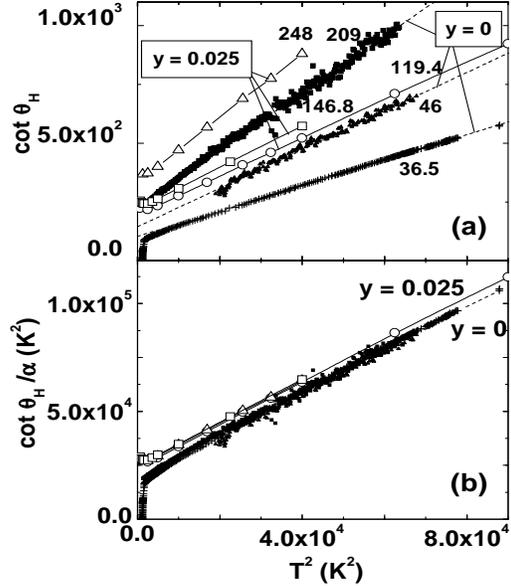, height=0.64\textwidth, width=0.5\textwidth}
\vspace*{-2.5cm}
\caption{$\cot{\Theta_H}$ at 8 T (a) and $\cot{\Theta_H}/\alpha$ (b) 
as a function of $T^2$ for $y=0$ (full points, dashed lines), and
$y=0.025$ (open points, solid lines). The lines are fits to the relation
$\cot \Theta_H = \alpha T^2 + C$. In (a) the data are labeled with the
value of the residual resistivity $\rho_0$ (in ${\mu \Omega}\,$cm).}
\end{figure}

The proportionality of $C$ and $\alpha$ persists for all values of $y$,
i.e. {\em strain affects these two parameters in the same way}.
The dependence of $C/\alpha$ on $y$ is shown on Fig.~4a.
Since $\alpha$ is $y$--independent, it follows that $C = C_0 + C_1y$.
The straight line fitted to the data gives 
$C_0 /{\alpha}= 1.8 \times 10^4$ K$^2$, and $C_1/{\alpha} = 2.6 \times
10^5$ K$^2$.

\begin{figure}[ht]
\vspace*{-0.6cm}
\epsfig{file=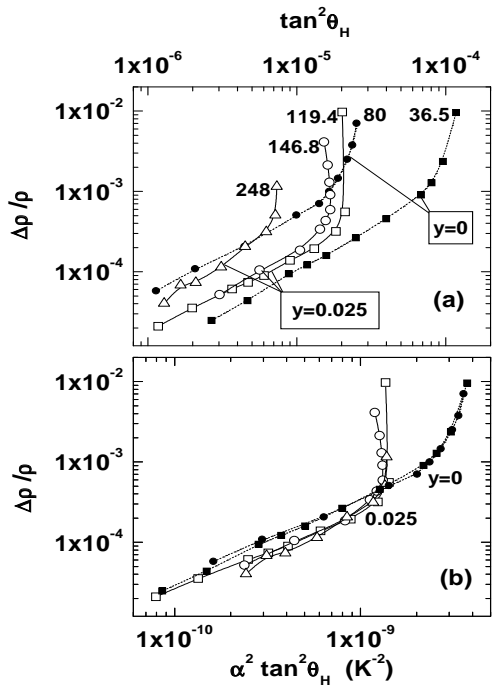, height=0.64\textwidth, width=0.5\textwidth}
\vspace*{-2.1cm}
\caption{Log--log plot of the orbital magnetoresistance versus $\tan^2
\Theta_H$ (a)
and versus $\alpha^2 \tan^2 \Theta_H$ (b). The data are for two groups
of films:
$y=0$ (full symbols, dashed lines), and $y=0.025$ (open symbols,
solid lines). All lines are guides to the eye. In (a) the data are
labeled with
the value of $\rho_0$ for each film. }
\end{figure}

We can now summarize the behavior of the Hall angle by stating that
although $\cot{\Theta_H}$ depends on both strain and
impurity, $(\cot{\Theta_H})/\alpha$, which is equal to $T^2 + C/\alpha$,
depends on $y$ only. This is illustrated on Fig.~5.
Fig.~5a shows $\cot{\Theta_H}$ vs. $T^2$ for six specimens with two
values of $y$. On Fig.~5b the same data are shown as
$(\cot{\Theta_H})/\alpha$ vs. $T^2$, and are seen to collapse to two
curves, one for each value of $y$. 

A similar situation arises for the magnetoresistance. We can rewrite
the relationship between the OMR and the Hall angle as
$\Delta \rho/\rho = (\zeta/\alpha^2)(\alpha^2 \tan^2{\Theta_H})$. The
experiment shows that {\em the OMR does not depend on strain}. Since
$\alpha \tan{\Theta_H}$ also depends only on $y$, it follows that this must be
the case also for $\zeta/\alpha^2$. Fig.~4b shows the variation of the
reciprocal of this quantity with $y$, which is seen to be a straight
line with intercept $2.75 \times 10^{-6}$ K$^{-4}$ and slope
$6.54 \times 10^{-5}$ K$^{-4}$.
The point for the single crystal of Ref. \cite{harris} is seen to be
consistent with this line. Since $\alpha$ is $y$--independent, this is
also the $y$--dependence of $1/\zeta$.

Fig.~6a shows the OMR plotted against $\tan^2{\Theta_H}$ for five films
with two values of $y$. The high--temperature parts of the data,
between 70K and 200K, correspond to low OMR and low $\tan{\Theta_H}$.
They fall on straight lines with equal slopes, indicating that the OMR
and $\tan^2{\Theta_H}$ are indeed proportional, as also shown in Ref.
\cite{harris}. Near 70K the data for all films deviate from the straight
lines,
independent of $y$ and $\rho_s$ \cite{fedor,malin}. Fig.~6b shows the
same data plotted against ${\alpha^2}\tan^2{\Theta_H}$, where they are
seen to collapse to two lines, one for each value of $y$.

We can now compare our results with the existing theoretical models
which attempt to explain the anomalous normal state
\cite{harris,and,col,car,stoj,hlub,ioffe,zhele,sand,varma,hlubina2}.
First, the non-Fermi-liquid models assume the existence of
two different relaxation rates at all points of the Fermi surface \cite
{harris,and,col}. In these models the
ratio $\zeta$ is constant and should not be affected by impurities.
Our results are not compatible with this expectation.
On the other hand, the Fermi--liquid models are based on the
assumption that strong anisotropy of the relaxation rates around
the Fermi surface leads to the anomalies \cite{car,stoj,hlub,ioffe,zhele,sand}.
The details of these models vary, but
generally it would be expected that the ratio $\zeta$ would depend on
impurities. While this seems to agree with the experiment,
the strong effect of strain on the Hall effect and lack of any strain
effect on the OMR is not expected. The simultaneous proportional change
of $C$ and $\alpha$ could be ascribed to
the influence of strain on cyclotron frequency, but the insensitivity
of the OMR to strain is then puzzling. It is conceivable that the
cyclotron frequency remains constant and strain
affects both the inelastic and elastic Hall scattering rates in such a
way that they cancel in the OMR. This seems improbable.
Other experiments, including the infrared Hall effect and the angle--resolved 
photoemission, give evidence of non--Fermi--liquid behavior \cite{cerne,valla}.
Motivated by these results, Varma and Abrahams proposed a new model
based on angle--independent marginal--Fermi--liquid inelastic scattering
coupled with angle--dependent elastic scattering \cite{varma,hlubina2}.
It remains to be seen whether it leeds to better description of 
the dc--transport properties.

We conclude with the modifications and additions to the equations of
the first paragraph: $\rho_0 = \rho_s + \rho_y$, $C = C_0 + C_1 y$,
$\cot{\Theta_H}/{\alpha} = T^2 + C/{\alpha} = f(y)$ and
$\Delta \rho / \rho = (\zeta /\alpha^2)({\alpha^2}{\tan^2{\Theta_H}})$,
where each term in brackets is a function of $y$ only.

We would like to thank M. Gershenson and S-W. Cheong for their cooperation 
and sharing of laboratory facilities. We thank Piers Coleman and 
Andrew Millis for
helpful discussions. We also thank T. E. Madey for providing the
facilities and helping with the AFM interpretation, and C. L. Chien 
for the 4--cycle x--ray diffractometer measurements. 
This work was supported by the Polish Committee
for Scientific Research, KBN, under grants 2 P03B 09414 and 7 T08A 01115,
by the Naval Research Laboratory, by the Rutgers Research Council,
and by the Eppley Foundation for Research.

\end{document}